# Absorption enhancement in photonic crystal thin films by pseudo disordered perturbations

Romain Peretti, Guillaume Gomard[1], Loïc Lalouat, Christian Seassal and Emmanuel Drouard[*]

Institut des Nanotechnologies de Lyon (INL), Université de Lyon, UMR 5270, CNRS-INSA-ECL-UCBL, Ecole Centrale de Lyon, 36 Avenue Guy de Collongue, 69134 Ecully Cedex, France

[1] Present address: Light Technology Institute (LTI), Karlsruhe Institute of Technology (KIT), 76131 Karlsruhe, Germany

[*] Corresponding author: emmanuel.drouard@ec-lyon.fr

**ABSTRACT**

The effects resulting from the introduction of a controlled perturbation in a single pattern membrane on its absorption are first studied and then analyzed on the basis of band folding considerations. The interest of this approach for photovoltaic applications is finally demonstrated by overcoming the integrated absorption of an optimized single pattern membrane through the introduction of a proper pseudo disordered perturbation.

**INTRODUCTION**

Photonic nanostructures have been intensely investigated in the framework of thin and ultra-thin film solar cells for light management purposes in order to improve the light collection, and enhance the absorption close to the energy bandgap of the active material through light-trapping schemes. In this context, the concept of absorbing photonic crystal (PhC) has emerged [1]. It is based on the patterning of the active layer itself, in which nanoholes are periodically drilled. Theoretical and experimental studies have demonstrated that such a configuration, if properly designed, could be beneficial for the absorption over the whole spectral range considered [2]. Thus, at short wavelengths, the PhC patterning leads to a significant decrease of the reflection losses while at large wavelengths, new absorption peaks appear and result from the coupling of the incident light with pseudo guided modes. Beyond those remarkable features, it turned out that due to the limited number of pseudo guided modes supported by this periodical structure, it is highly challenging to avoid large dips in the resulting absorption spectrum close to the bandgap. Motivated by the prospect of further enhancing the absorption in this spectral region and improving the angular and polarization tolerance of the patterned layers, several research groups have recently studied the introduction of structural disorder within those PhC membranes. The latter, which is most often position disorder, falls into three categories: multiperiodic patterns, correlated-disordered configurations and random structures. Using a controlled amount of disorder, absorption enhancements with regards to a single pattern PhC could be reported, which was attributed to more favorable coupling conditions and to the introduction of new absorption peaks [3-5].

In this paper, we propose to analyze the origin and the properties of the added modes so as to facilitate the engineering of such perturbed structures.

**OPTIMIZATION OF THE INTEGRATED ABSORPTION IN SINGLE PATTERN PHC MEMBRANES**

Let us first consider a single pattern membrane possessing an isolated absorption peak in its low-absorbing region. This resonance is supposed to result from the coupling to a slow Bloch mode (SBM) which is pseudo guided in the membrane. Using the time domain coupled mode theory, it can be shown that a maximal peak absorption is reached at the resonant wavelength of the mode when the absorption losses of the mode equal the external ones [2]. This situation is known as the critical coupling conditions, and can lead to absorptions up to 100% at a single wavelength if the absorption of the peak considered is added to the absorption of Fabry-Pérot and/or surrounding SBM resonances [6,7]. The optimal operation regime is nevertheless different if one wants to optimize the integrated absorption of the isolated peak considered, which is the relevant quantity for photovoltaic applications. Indeed, its value depends not only on the peak absorption, but also on the width of the absorption peak. The latter, which can be related to the quality factor ($Q$) of the mode, increases together with its external losses. Consequently, it appears that the integrated absorption of an isolated peak is maximized when the external losses are dominating over the absorption ones (over-coupling regime) [8].

Thus, the addition of disorder in a regular lattice potentially enables to enhance the integrated absorption of the membrane via two mechanisms: the filling of the low-absorbing region with new absorption peaks thanks to the coupling of the incident light into additional SBM, and the improvement of the coupling conditions to maximize the integrated absorption of each mode. Those two mechanisms are reviewed in more details in the following sections.

**EFFECTS OF THE INTRODUCTION OF A PERTURBATION IN MEMBRANES**

<u>Methodology and structures studied</u>

The absorption of the structures studied, calculated for a normal incidence of the light, is obtained using Finite-difference time-domain (FDTD) simulations by taking into account the chromatic dispersion of the optical indices of a-Si:H (see [9] for details). The integrated absorption of the structures (under AM1.5G illumination) and the corresponding short circuit current density $J_{sc}$ (assuming an internal quantum efficiency of 1) can thus be easily derived. To analyze the modes involved in the absorption, their $Q$ factors are estimated using a harmonic inversion algorithm (Harminv [10]). Those quality factors are compared to the critical quality factor $Q_c$ [9] which is obtained when the external losses equal the intrinsic (absorption) losses. It is used as a benchmark for identifying the various coupling regimes, especially the over-coupling regime where $Q$ is lower than $Q_c$.

The first configuration investigated in this paper, used as a reference, is based on a feasible structure and consists in a 195 nm thin layer of hydrogenated amorphous silicon (a-Si:H) on a glass substrate. The absorbing layer is fully etched by cylindrical nano-holes arranged according to a square lattice (corresponding to a $C_{4V}$ point group), as depicted on Figure 1.a (left). For the sake of clarity, the geometrical parameters (period $a = 220$ nm and surfacic air filling fraction $ff = 50\%$) of this single pattern (SP) membrane are chosen so that only one absorption peak is present in the wavelength range where resonant modes ($Q>10$) are needed, namely above around 600 nm, where the extinction coefficient of a-Si:H is below 0.17 [9].

The corresponding absorption spectrum is reported on Figure 1.a (right), together with the one of an equivalent unpatterned membrane for comparison. Three modes are identified by the modal analysis: two of them are anti-symmetric with respect to the (*xy*) plane of the membrane, and therefore cannot couple to a plane wave under normal incidence (i.e. in Γ) [9,11], and one of them is symmetric and generates an absorption peak at 666 nm as its coupling with the incident light is allowed. It can be noted that the over-coupling regime is barely reached, since its Q factor is almost equal to $Q_c$.

To investigate the effects of disorder, a second configuration, referred to as "Multiple Pattern" (MP), will be simulated. It corresponds to a simple perturbation scenario as only the position of some holes is changed while conserving the overall $C_{4v}$ symmetry and *ff*. Considering a supercell of size *3a*x*3a*, this can be achieved by moving the four holes located at the corners of the supercell towards its center, as shown on Figure 1.b. This symmetric multiple pattern configuration will be used to derive general conclusions on the effects of disorder, which will be then extended to any non-symmetric and space-correlated disorder structures. We express the magnitude of the opto-geometric perturbation (denoted *M*) applied to a supercell of period *Na* as:

$$M(\lambda) = \Delta n(\lambda) \times \left(\frac{1}{(Na)^2}\right) \times \iint_{perturbation} |dS| = \Delta n(\lambda) \Delta S$$

(1)

where the first term is the refractive index difference between a-Si:H and air, and the second one is the infinitesimal displacement area induced by the perturbation and integrated along the *x* and *y* directions. Since the refractive index difference is similar for the different perturbations, the latter can be simply expressed by the magnitude of surface perturbation ΔS (in square period units) which does not depend on the wavelength considered. In our case, ΔS will be limited by the fact that the overlap of holes is not allowed in order to keep the same *ff* and facilitate the analysis of our results.

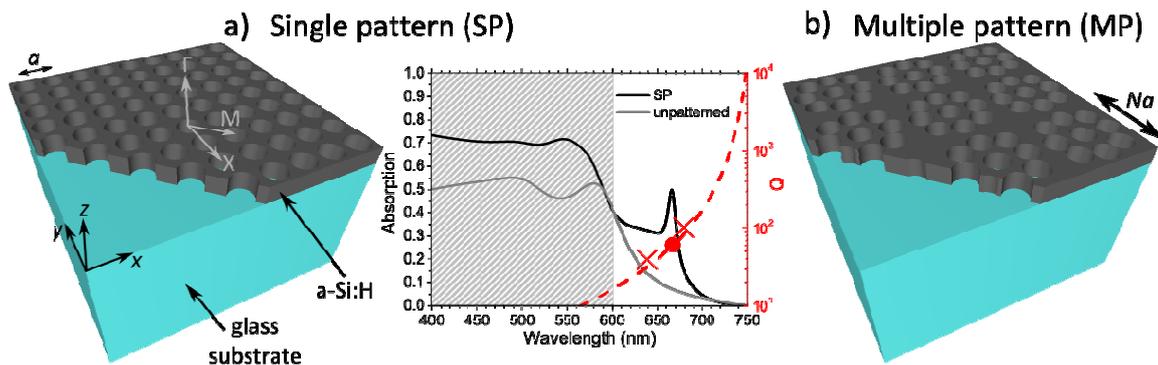

**Figure 1.** a) Schematic representation of the single pattern membrane (left) and absorption spectra of the unpatterned and single pattern membranes (left axis) in correspondence with the calculated $Q_c$ (red dashed line, right axis) and measured Q of the symmetric (red dot) and anti-symmetric (red crosses) modes (right). The crystallographic directions are also reported for a further analysis b) Schematic representation of the multiple pattern membrane (*N*=3).

**Effects of the introduction of controlled disorder in symmetric patterned membranes**

This section aims at emphasizing the consequences of introducing a controlled amount of disorder inside the SP membrane leading to the MP configuration presented on Figure 1.b ($N = 3$). We will focus on the wavelength range of interest between 600 and 750 nm.

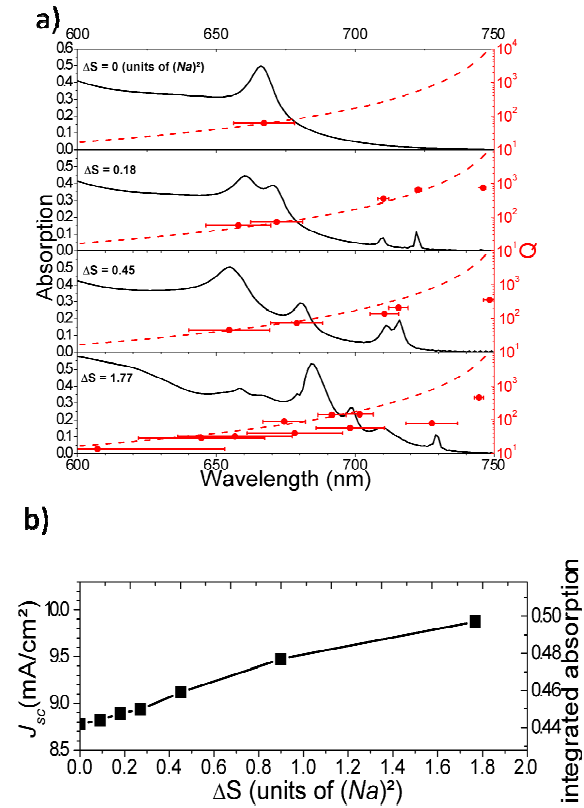

**Figure 2.** a) Evolution of the absorption spectrum and of the $Q$ factors of the modes with $\Delta S$ in the MP structure ($Q_c$ depicted with dashed line. b) Influence of the perturbation magnitude on its $J_{sc}$ and integrated absorption.

As soon as a small perturbation is introduced, new modes appear as highlighted on **Figure 2.**a. Globally, they are spectrally separated from the mode at 666 nm, and therefore only weakly perturbing the original peak. The main consequence is the creation of additional absorption peaks. Both phenomena can be translated into an enhancement of the integrated absorption (see **Figure 2** b). The latter keeps increasing as $\Delta S$ tends to its highest limit, gaining up to +5.5% in absolute value with regards to the SP ($\Delta S=0$) configuration. A closer look at the $Q$ factors of the new modes shows that they are progressively decreased when $\Delta S$ is strengthened, sometimes far below their corresponding $Q_c$ (over-coupling regime), with the already mentioned consequences of the broadening of the peaks and an improvement of the overall integrated absorption [8]. Finally, as the resonant wavelengths of the added modes are also sensitive to the perturbation, this result in an overlap of the modes and less dips in the spectrum.

**Increase of the spectral density of modes through band folding mechanisms**

The creation of additional modes which have the potential to couple to an incident plane wave can be qualitatively explained by considering the band folding mechanisms occurring within the reduced Brillouin zone (RBZ) of the structures studied and assuming very low perturbations ($\Delta S \rightarrow 0$). This situation is depicted on **Figure 3**.a. As can be seen on this simplified schematic, as the period of the MP supercell is 3 times higher than for the unperturbed cell (in real space), several high-symmetry points corresponding to the MP are introduced in the first RBZ of the SP structure. Taking the example of the $\Gamma_{MP}$ point represented by the black dot, the folding mechanism may allow to couple modes in vertical direction ($\Gamma_{MP}$) which were not located at $\Gamma_{SP}$. Two folding schemes, summarized on **Figure 3 .b**, can be distinguished: either the folded branches cross at the same frequency giving rise to a degenerated mode which can couple to a normal plane wave [9,11], or cross at different frequencies (anti-crossing configuration) in which case the added modes in $\Gamma_{MP}$ will not be able to couple with the normal incident light. It follows that only the former situation will generate new absorption peaks. It should be also precised that the present description gives information on the number of modes introduced, but not on their frequencies.

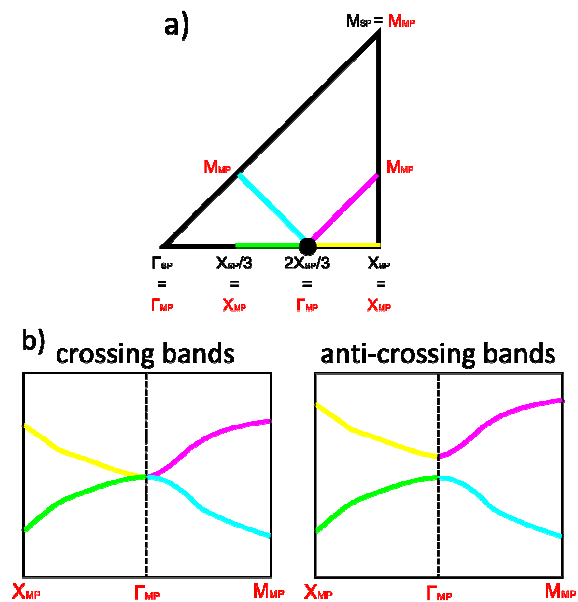

**Figure 3.** a) First RBZ of the SP structure (in black, labeled "SP") depicted with some high-symmetry $k$ points of the MP supercell (in red, labeled "MP"). b) Band folding corresponding to a crossing (left) or anti-crossing (right) situation.

In addition, a rigorous analysis of this symmetric MP structure in its reciprocal space, as developed in [9] for modes that can couple under normal incidence, enabled to conclude that the density of modes is increased by a factor of $N^2$ for supercells whose unit cell area is $N^2$ times larger than that of the SP structure (provided that the spectral range of interest is broad enough). Theoretically, this can potentially lead to $N^2$ times more peaks in the absorption spectrum, although it should be precised that the actual number of peaks contributing to a real increase of the absorption is lower as they should be located in the low-absorbing spectral region and present favourable coupling conditions. This demonstration can be generalized to any space-correlated disorder structures (with a supercell period $N$) such as non-symmetric MP configurations. As the original symmetry is broken, the folded modes are split into two as they are no longer degenerated. Besides, the other modes which were uncoupled for symmetry reasons can now be coupled under normal incidence. As they are twice as numerous as symmetric modes, the overall number of modes which can be coupled with such perturbations is $4N^2$ times the number of modes in the SP case.

**Case study: absorption enhancement in a pseudo disordered membrane**

As an illustration of the previous sections, we studied the effects of a pseudo disordered perturbation for which the holes are moved in their own subcell according to a normal distribution. This perturbation was introduced in an optimized SP membrane ($a$ = 345 nm, $ff$ = 50%) [9]. As evidenced on Figure 4, the pseudo disordered $3a$x$3a$ membrane presents not only more absorption peaks above 600 nm than the SP membrane but also broader ones, resulting in an absolute integrated absorption increase of 2%. This value was only obtained for $\Delta S = 0.6(Na)^2$, suggesting the existence of an optimum value of the perturbation as underlined in other studies [3].

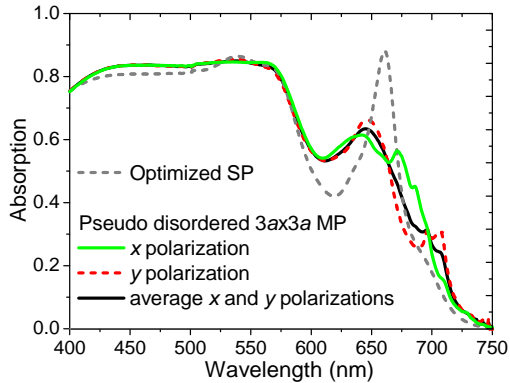

**Figure 4:** Absorption increase in an optimized SP structure via a pseudo disordered perturbation.

**CONCLUSIONS**

The absorption of a single patterned membrane can be overcome by the introduction of a controlled amount of position disorder thanks to the broadening of the absorption peaks and the increase of the density of modes which directly depends on the size of the supercell considered.

**ACKNOWLEDGMENTS**

This work was supported by the European Community (FP7 project PhotoNvoltaics N°309127) and by Orange Labs Networks (contract 0050012310-A09221).